\documentclass[10pt,letterpaper]{article}
\usepackage[top=0.85in,left=2.75in,footskip=0.75in,marginparwidth=2in]{geometry}

\usepackage[utf8]{inputenc}

\usepackage{cite}

\usepackage{nameref,hyperref}

\usepackage{microtype}
\DisableLigatures[f]{encoding = *, family = * }

\raggedright
\usepackage{changepage}

\usepackage[aboveskip=1pt,labelfont=bf,labelsep=period,singlelinecheck=off]{caption}

\makeatletter
\renewcommand{\@biblabel}[1]{\quad#1.}
\makeatother

\usepackage{lastpage,fancyhdr,graphicx}
\usepackage{epstopdf}
\fancyheadoffset[L]{2.25in}
\fancyfootoffset[L]{2.25in}

\usepackage{color}

\definecolor{Gray}{gray}{.25}

\usepackage{graphicx}

\usepackage{sidecap}

\usepackage{wrapfig}
\usepackage[pscoord]{eso-pic}
\usepackage[fulladjust]{marginnote}

\usepackage[english]{babel}  
\usepackage{amsmath}  
\usepackage{amsfonts}  
\usepackage{amssymb}  
\usepackage{wasysym}  
\usepackage{bm}  
\usepackage{array}  
\usepackage{xr}  
\usepackage{verbatim} 
\usepackage{float} 

\usepackage{soul}
\usepackage{algorithm}
\usepackage{algorithmic}
\usepackage{color}
\usepackage{underscore}
\reversemarginpar

\begin{document}
\vspace*{0.35in}

\begin{flushleft}
{\Large
\textbf\newline{Deep Manifold Learning for Dynamic MR Imaging}
}
\newline
\\
Ziwen Ke \textsuperscript{1, 2, {†}},
Zhuo-Xu Cui \textsuperscript{3, {†}},
Wenqi Huang \textsuperscript{1, 2},
Jing Cheng \textsuperscript{3},
Sen Jia \textsuperscript{3},
Haifeng Wang \textsuperscript{3},
Xin Liu \textsuperscript{3},
Hairong Zheng \textsuperscript{3},
Leslie Ying \textsuperscript{4},
Yanjie Zhu \textsuperscript{3},
Dong Liang \textsuperscript{1, {*}}
\\
\bigskip
\bf{1} Research Center for Medical AI, Shenzhen Institutes of Advanced Technology, Chinese Academy of Sciences, Shenzhen, China
\\
\bf{2} Shenzhen College of Advanced Technology, University of Chinese Academy of Sciences, Shenzhen, China
\\
\bf{3} Paul C. Lauterbur Research Center for Biomedical Imaging, Shenzhen Institutes of Advanced Technology, Chinese Academy of Sciences, Shenzhen, China
\\
\bf{4} Department of Biomedical Engineering and the Department of Electrical Engineering, The State University of New York, Buffalo, NY, USA
\\
\bigskip

\textbf{{†}}: These authors contributed equally to this work.

\textbf{*} Corresponding author:

\indent\indent
\begin{tabular}{>{\bfseries}rl}
	Name	& 	Dong Liang												\\
	Department	&  Research Center for Medical AI													\\
	Institute	& Shenzhen Institutes of Advanced Technology, Chinese Academy of Sciences														\\
	Address 	& 1068 Xueyuan Avenue, Shenzhen University Town														\\
	& Shenzhen														\\
	& China														\\
	E-mail		& dong.liang@siat.ac.cn											\\
\end{tabular}

\end{flushleft}

\section*{Abstract}

\textbf{Purpose}: To develop a deep learning method on a nonlinear manifold to explore the temporal redundancy of dynamic signals to reconstruct cardiac MRI data from highly undersampled measurements.

\textbf{Methods}: Cardiac MR image reconstruction is modeled as general compressed sensing (CS) based optimization on a low-rank tensor manifold. The nonlinear manifold is designed to characterize the temporal correlation of dynamic signals. Iterative procedures can be obtained by solving the optimization model on the manifold, including gradient calculation, projection of the gradient to tangent space, and retraction of the tangent space to the manifold. The iterative procedures on the manifold are unrolled to a neural network, dubbed as Manifold-Net. The Manifold-Net is trained using in vivo data with a retrospective electrocardiogram (ECG)-gated segmented bSSFP sequence.

\textbf{Results}: Experimental results at high accelerations demonstrate that the proposed method can obtain improved reconstruction compared with a compressed sensing (CS) method k-t SLR and two state-of-the-art deep learning-based methods, DC-CNN and CRNN.

\textbf{Conclusion}: This work represents the first study unrolling the optimization on manifolds into neural networks. Specifically, the designed low-rank manifold provides a new technical route for applying low-rank priors in dynamic MR imaging.


\section{Introduction}

In 2019, ischaemic heart disease was the 1st leading cause of death worldwide, responsible for 16\% of the world’s total deaths \cite{who2020the}. Many imaging modalities, namely, magnetic resonance imaging (MRI), computed tomography (CT), and ultrasonic, are employed for heart diagnosis. The key to the diagnosis is whether the imaging methods can capture the cardiac structure and dynamic motion. Of these modalities, not only is MRI free from ionizing radiation and the use of radioactive tracers, but also with unparalleled soft-tissue contrast and high spatial resolution. However, MRI has yet to become a routine part of clinical workup in cardiac patients \cite{rajiah2014cardiovascular}, primarily due to the limitation: speed. The acquisition time would increase significantly if the more excellent spatial resolution, volume coverage, and motion-corrupted acquisitions are needed. To overcome this limitation, acceleration methods from highly under-sampled k-space measurements have been intensely studied for nearly three decades.

Accelerated dynamic MRI is unique because of an extra time dimension. Exploration of the Spatio-temporal correlation of dynamic signals plays an essential role in maximizing acceleration. k-t accelerated imaging \cite{madore1999unaliasing, tsao2003k, huang2005k, pedersen2009k} is one of the earliest acceleration methods that reduce the reconstruction problem to a mathematical estimation via a linear combination of acquired points in k-t space. To capture and utilize global similarities for ensuring more accurate reconstruction, compressed sensing (CS) \cite{donoho2006compressed, lustig2007sparse, otazo2010Combination} has been utilized. According to CS, incoherent artifacts from random undersampling can be removed in some transform (sparse basis) domains by nonlinear reconstruction. From the initial fixed basis \cite{lustig2006kt, jung2007improved, liang2012k}, to the sparse adaptive basis \cite{caballero2014dictionary, wang2013compressed, nakarmi2015dynamic}, and then to the recent use of a neural network to learn the sparse basis \cite{schlemper2017deep, qin2018convolutional, wang2019dimension, biswas2019dynamic}, the sparse prior in CS is more and more intelligent with continuously improved reconstruction performance.

Recently, manifold learning \cite{poddar2015dynamic, nakarmi2017kernel, shetty2019bi, ahmed2020dynamic} has achieved some success in accelerated cardiac MRI. The core of manifold learning is an assumption that the dynamic signals in high dimensional space are adjacent points on smooth and low dimensional manifolds. The construction of low-dimensional manifolds is a topic of ongoing research. For example, in SToRM \cite{poddar2015dynamic}, a graph Laplacian matrix is established from a navigator acquisition scheme to determine the structure of the manifold. KLR \cite{nakarmi2017kernel} used kernel principal component analysis (KPCA) \cite{mika1998kernel} to learn the manifold described by the principal components of the feature space. BiLMDM \cite{shetty2019bi} extracted a set of landmark points from data by a bi-linear model to learn a latent manifold geometry. These methods posed the manifold smoothness prior as a regularization in a CS optimization problem and performed iterative optimization in the linear European space to obtain reconstructed results. Although these works make generous contributions to dynamic MRI, the following issues still need to be addressed: 1) the manifold smoothness prior is located in nonlinear manifolds, while corresponding iterative optimization is not performed along with the nonlinear structure of the manifold; 2) Manifold regularization, as an extension of CS, cannot escape the essential iterative reconstruction process, and cannot avoid tedious reconstruction time and parameter selection.

Inspired by Riemannian optimization \cite{absil2009optimization, kressner2014low}, we propose a deep manifold learning for dynamic MRI in this paper. In particular, a low-rank tensor manifold is designed to characterize the strong temporal correlation of dynamic signals. Dynamic MR reconstruction is modeled as a general CS-based optimization on the manifold. Riemannian optimization on the manifold is used to solve this problem, and iterative procedures can be obtained. To avoid the disadvantages of long iterative solution time and difficult parameter selection, these iterative procedures are unrolled into a neural network, dubbed as Manifold-Net. Extensive experiments on in vivo MRI data illustrate noteworthy improvements of the proposed Manifold-Net over state-of-the-art reconstruction techniques. This work represents the first study unrolling the optimization on manifolds into neural networks. Besides, the designed low-rank manifold provides a new technical route for applying low-rank priors in dynamic MR imaging.

The rest of this paper is organized as follows. Section II and III provide the recall of Riemannian optimization theory and introduces the proposed methods. Section IV summarizes the experimental details and the results to demonstrate the effectiveness of the proposed method, while the discussion and conclusions are presented in Section IV and V, respectively.

\section{Theory}

Tensor completion \cite{liu2012tensor} fills in missing entries of a partially known tensor with a low-rank constraint. Riemannian optimization techniques on a manifold of tensors of fixed rank have been mathematically and experimentally proved feasible to solve the tensor completion problem \cite{kressner2014low}—this section recall Riemann optimization theory. Detailed introduction of theoretical derivation is beyond the scope of this paper, so we focus on its three important iterative steps: gradient calculation, projection of the gradient to tangent space, and retraction of the tangent space to the manifold. Throughout this section, we follow the notation in \cite{kressner2014low}.

\subsection{Preliminaries on Tensors}
\begin{itemize}
	\item{$The\ ith\ mode\ matricization$:}
	
	The $i$th mode matricization of a tensor $\bm{x}\in \mathbb{C}^{n_1\times\dots\times n_d}$ is a rearrangement of the entries of $\bm{x}$ into the matrix $\bm{x}_{(i)}$, such that the $i$th mode becomes the row index and all other $(d-1)$ modes become column indices, in lexicographical order.
	\begin{equation}
	\label{mode}
	\bm{x}_{(i)} \in \mathbb{C}^{n_i \times \Pi_{j\neq i}n_j}
	\end{equation}
	
	\item{$The\ multilinear\ rank$:}
	
	The ranks of all the mode matricizations yield the multilinear rank tuple $r$ of $\bm{x}$:
	\begin{equation}
	\label{rank}
	\text{rank}(\bm{x})=(\text{rank}(\bm{x}_{(1)}), \text{rank}(\bm{x}_{(2)}), \dots, \text{rank}(\bm{x}_{(d)}))
	\end{equation}
	
	\item{$The\ ith\ mode\ product$:}
	
	The $i$th mode product of $\bm{x}$ multiplied with a matrix $M\in \mathbb{C}^{m\times n_i}$ is defined as
	\begin{equation}
	\label{mode_product}
	\begin{aligned}
	&\bm{y}=\bm{x}\times_i M \Leftrightarrow \bm{y}_{(i)}=M\bm{x}_{(i)}, \\
	&\bm{y}\in \mathbb{C}^{n_1 \times \dots \times n_{n-1}\times m \times n_{i+1}\times \dots \times n_d}
	\end{aligned}
	\end{equation}
	
	\item{$Tucker\ decomposition$:}
	
	Any tensor of multilinear rank of $r=(r_1, r_2, \dots, r_d)$ can be represented in the so-called $Tucker\ decomposition$
	\begin{equation}
	\label{tuker_decom}
	\bm{x}=C\times_1 U_1 \times_2 U_2 \cdot\cdot\cdot \times_d U_d=C \mathop{\times}\limits_{i=1}^{d} U_i
	\end{equation}
	with the $core\ tensor$ $C \in \mathbb{C}^{r_1\times\dots\times r_d}$, and the $basis\ matrices$ $U_i\in \mathbb{C}^{n_i\times r_i}$. Without loss of generality, all $U_i$ are orthonormal: $U^T_iU_i=I_{r_i}$.
	
	\item{$Higher\ order\ singular\ value\ decomposition$:}
	
	The truncation of a tensor $\bm{x}$ to a fixed rank $r$ can be obtained by the higher order singular value decomposition (HOSVD) \cite{de2000multilinear}, denoted as $\text{P}_\text{r}^{\text{HO}}$. The HOSVD procedure can be described by the successive application of best rank-$r_i$ approximations $\text{P}_{r_i}^{i}$ in each mode $i=1,\dots, d$:
	\begin{equation}
	\label{HOSVD}
	\text{P}_\text{r}^{\text{HO}}:\ \mathbb{C}^{n_1\times\cdot\cdot\cdot\times n_d} \to \mathcal\bm{{M}_r},\ \ \bm{x}\ \mapsto \text{P}_{r_d}^{d}\circ\cdots\circ \text{P}_{r_1}^{1}\bm{x}.
	\end{equation}
	Each individual projection can be computed by a truncated SVD as follows. Let $U_Y$ contain the $r_i$ dominant left singular vectors of the $i$th matricization $Y_{(i)}$ of a given tensor $\bm{y}$. Then the tensor resulting from the projection $\tilde{\bm{y}}=\text{P}_{r_i}^{i}\bm{y}$ is given in terms of its matricization as $\tilde{Y}_{(i)}=U_YU_Y^TY_{(i)}$. The HOSVD does inherit the smoothness of low-rank matrix approximations \cite{chern2001smoothness}.
\end{itemize}

\subsection{Manifold setting}

The set $\mathcal\bm{{M}_r}$ of tensors of fixed rank $r=(r_1, r_2, \dots, r_d)$ forms a smooth embedded submanifold of $\mathbb{C}^{n_1\times\cdots\times n_d}$ \cite{uschmajew2013geometry}. By counting the degrees of freedom in (\ref{tuker_decom}), it follows that the dimension of $\mathcal\bm{{M}_r}$ is given by
\begin{equation}
\label{dimension}
\text{dim}(\mathcal\bm{{M}_r})=\prod_{j=1}^d r_j+\sum_{i=1}^{d}r_in_i-r_i^2
\end{equation}

According to \cite{koch2010dynamical}, the $tangent\ space$ of $\mathcal\bm{{M}_r}$ at $\bm{x}=C\times_1 U_1 \times_2 U_2 \cdots \times_d U_d$ can be parametrized as
\begin{equation}
\label{tangent_space}
T_{\bm{x}}\mathcal\bm{{M}_r}=\left\{ G\mathop{\times}\limits_{i=1}^{d}U_i+\sum_{i=1}^{d}C\times_iV_i\mathop{\times}\limits_{j\neq i}U_j|V_i^TU_i=0\right\}
\end{equation}
where $G\in \mathbb{C}^{r_1\times\cdots\times r_d}$ and $V_i\in \mathbb{C}^{n_i\times r_i}$ are the free parameters. Furthermore, the orthogonal projection of a tensor $A\in \mathbb{C}^{n_1\times\cdots\times n_d}$ onto $T_{\bm{x}}\mathcal\bm{{M}_r}$ is given by
\begin{equation}
\label{orthogonal_proj}
\begin{aligned}
\text{P}_{T_{\bm{x}}\mathcal\bm{{M}_r}}&:\ \mathbb{C}^{n_1\times\cdots\times n_d}\rightarrow T_{\bm{x}}\mathcal\bm{{M}_r}\\
A&\mapsto\left(A\mathop{\times}\limits_{j=1}^{d}U_j^T\right)\mathop{\times}\limits_{i=1}^{d}U_i\\&+\sum_{i=1}^{d}C\times_i\left(\text{P}\frac{1}{U_i}\left[A\mathop{\times}\limits_{j\neq i}U_j^T\right]_{(i)}C_{(i)}^\dagger\right)\mathop{\times}\limits_{k\neq i}U_k
\end{aligned}
\end{equation}
Here $C_{(i)}^\dagger$ denotes the pseudo-inverse of $C_{(i)}$. Note that $C_{(i)}$ has been full row rank and hence $C_{(i)}^\dagger=C_{(i)}^T(C_{(i)}C_{(i)}^T)^{-1}$. We use $\text{P}\frac{1}{U_i}:=I_{r_i}-U_iU_i^T$ to denote the orthogonal projection onto the orthogonal complement of span($U_i$).

\subsection{Riemanian Gradient and Retraction}

As a metric on $\mathcal\bm{{M}_r}$, the Euclidean metric from the embedded space induced by the inner product (\ref{mode_product}). Together with this metric, $\mathcal\bm{{M}_r}$ becomes a Riemannian manifold. This, in turn, allows us to define the Riemannian gradient of an objective function, which can be obtained from the projection of the Euclidean gradient into the tangent space. 

Let $f: \mathbb{C}^{n_1\times\cdot\cdot\cdot\times n_d}$ be a cost function with Euclidean gradient $\nabla f_{\bm{x}}$ at point $\bm{x}\in \mathcal\bm{{M}_r}$. The Riemannian gradient \cite{absil2009optimization} of $f: \mathcal\bm{{M}_r} \to \mathbb{C}$ is given by
\begin{equation}
\label{Riemannian_gradient}
\text{grad}\ f(\bm{x})=\text{P}_{T_{\bm{x}}\mathcal\bm{{M}_r}}(\nabla f_{\bm{x}})
\end{equation}
Let $\xi: = -\text{grad}\ f(\bm{x})$ be Riemannian gradient update.

Retraction $\text{R}: T_{\bm{x}}\mathcal\bm{{M}_r} \to \mathcal\bm{{M}_r}$ maps an element from the tangent space at $\bm{x}\in \mathcal\bm{{M}_r}$ to the manifold $\mathcal\bm{{M}_r}$. HOSVD is  chosen to implement the retraction:
\begin{equation}
\label{Retraction_HOSVD}
\text{R}: T_{\bm{x}}\mathcal\bm{{M}_r},\ \ \ (\bm{x},\ \xi) \mapsto \text{P}_\text{r}^{\text{HO}}(\bm{x}+\xi)
\end{equation}

\begin{figure}[htbp]
	\centerline{\includegraphics[width=0.8\linewidth]{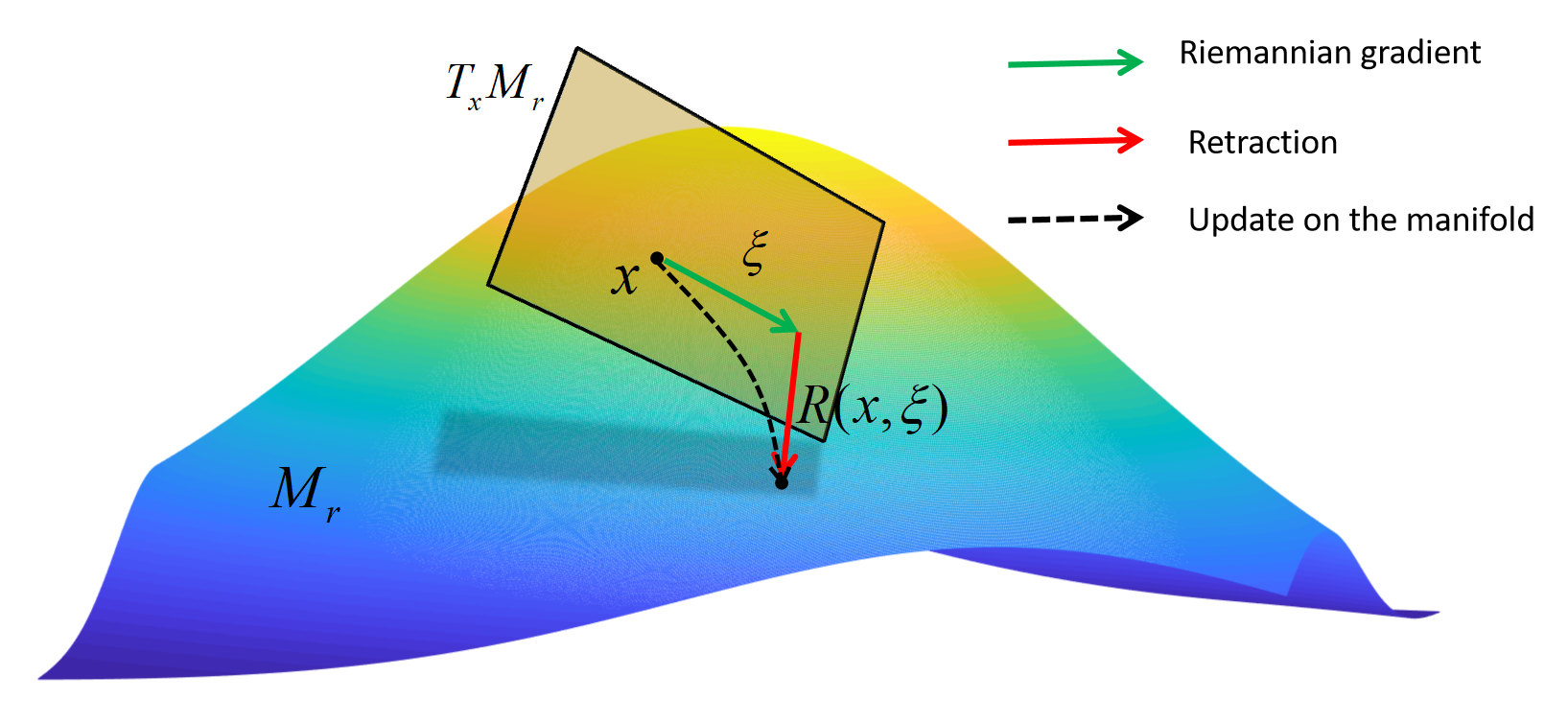}}
	\caption{Graphical representation of the concept of Riemannian gradient, retraction and update on the manifold within the framework of Riemannian optimization techniques. \label{riemannian_optimization}}
\end{figure}

A graphical depiction of these concepts is shown in Fig. \ref{riemannian_optimization}.

\section{Methods}
\subsection{Problem Formulation}
The reconstruction of dynamic MR images from under-sampled k-space data can be described as a tensor completion
problem. Namely, given only under-sampled k-space data, it is required to reconstruct the full-sampled MR image.
However, the completion process is ill-conditioned; that is, there are multiple solutions that are consistent with under-sampled data.  To reduce the range of solutions and stabilize the solution procedure, regularization is a common strategy. In consideration of the strong correlation between adjacent frames, the low-rank constraint is introduced to achieve regularization for dynamic MR imaging \cite{liang2007spatiotemporal, haldar2010spatiotemporal, lingala2011accelerated, zhao2012image, otazo2015low}.
This yields a nonlinear optimization problem as follows:
\begin{equation}
\label{SLR_model}
\min_{x\in \mathbb{C}^{N_xN_yN_t}}\frac{1}{2}\|A\bm{x}-\bm{y}\|^2+\lambda D(\bm{x}),~s.t.~\text{rank}(\bm{x})=r
\end{equation}
Where $A = PF$ is the encoding operator, $F$ denotes the Fourier transform, and $P$ denotes an under-sampling operator, $D$ is used to regularize each slice of $\bm{x}$. In particular, $D$ can be usually chosen as
$\sum_{t=1}^T\|W\bm{x}_t\|_1$, where $W$ denotes a certain sparsifying transform, such as wavelet, gradient mapping, etc.

For $r$-rank constrained optimization problem, iterative hard thresholding method \cite{foucart2013invitation} is the most common method:
\begin{equation}
\label{hard_svd}
\begin{aligned}
\bm{r}_{k+1}= &\bm{x}_k-\eta_k\left(A^{*}(A\bm{x}_k-\bm{y})+\lambda \nabla D(\bm{x}_k)\right)\\
\bm{x}_{k+1} =& \text{P}_\text{r}^{\text{HO}}(r_{k+1})
\end{aligned}
\end{equation}
where $\text{P}_\text{r}^{\text{HO}}$ denotes the higher-order singular value decomposition (HOSVD) operator (\ref{HOSVD}). The above hard algorithm can be regarded as gradient descent (GD) step compound with a hard thresholding mapping. GD is performed on Euclidean space
, so $r_{k+1}$ does not have rank $r$. An additional hard thresholding mapping is added to ensure that the iterate satisfies the $r$-rank constraint.

Since the GD step is only focused on seeking the direction towards an optimal solution on Euclidean space rather than low-rank constraint set and neglects the low-rank structure of seeking solutions, which usually leads to slow convergence of hard algorithms \cite{kressner2014low}. Furthermore, in this paper, we propose to design an effective optimization unrolled method for the problem (\ref{SLR_model}). Due to the limitation of computing capacity, the number of network layers of optimization unrolled method is usually chosen far less than the number of iterations of the traditional iterative algorithm. Therefore, the algorithms with slow convergence are not suitable for unrolling.

On the other hand, recent researches show that the set of tensors of fixed multilinear rank $r$ forms a smooth manifold \cite{kressner2014low}.
From this point of view, the low-rank constrained optimization problem (\ref{SLR_model}) reduces to an unconstrained optimization problem on a rank $r$ tensor formed manifold $\mathcal\bm{{M}_r}$:
\begin{equation}
\label{riemannian_model}\min_{\bm{x}\in \mathcal\bm{{M}_r}}f(\bm{x}):=\frac{1}{2}\|A\bm{x}-\bm{y}\|^2+\lambda D(\bm{x})
\end{equation}
By exploiting the manifold structure of $\mathcal\bm{{M}_r}$, it allows for the use of Riemannian optimization techniques.

\subsection{Riemannian Optimization}
To solve the Riemannian optimization problem (\ref{riemannian_model}), Riemannian GD (GD restricted to $\mathcal\bm{{M}_r}$) is one of the simplest and most effective schemes. The main difference from a hard algorithm is that every iterate of Riemannian GD always stays on the manifold $\mathcal\bm{{M}_r}$ and the calculations of gradient and iterative trajectory always follows the manifold structure itself.

Starting with the initial point $\bm{x}_0\in \mathcal\bm{{M}_r}$, there are following two main steps to executing the Riemannian GD: 1) calculate Riemannian gradient and 2) iterate on manifold along the direction of negative gradient. Because of the nonlinearity of manifolds, the gradient of $f$ is generalized as a point to the direction for greatest increase within the tangent space $T_{\bm{x}}\mathcal\bm{{M}_r}$, which in detail reads:
$\text{grad}f(x):=\text{P}_{T_{\bm{x}}\mathcal\bm{{M}_r}}(A^*(Ax-y)+\lambda \nabla D(x))$
where $\text{P}_{T_{\bm{x}}\mathcal\bm{{M}_r}}$ denotes a projection onto the tangent space $T_{\bm{x}}\mathcal\bm{{M}_r}$. The detailed calculation of Riemann gradient on manifold $M_r$ is shown in (\ref{orthogonal_proj}).

Unlike Euclidean space, one point moving in the direction of the negative Riemannian gradient is not guaranteed to always fall into manifold $\mathcal\bm{{M}_r}$. Hence, we will need the concept of a
retraction $R$ which maps the new iterate $\bm{x}+\eta\xi$ back to a point $R(\bm{x},\eta\xi)$ in manifold $\mathcal\bm{{M}_r}$. The detailed calculation of retraction mapping on manifold $\mathcal\bm{{M}_r}$ is shown in (\ref{Retraction_HOSVD}). The main iterative process of Riemannian GD  is shown in the Algorithm \ref{alg:1}.

\begin{algorithm}[htb]
	\caption{Riemannian GD for problem (\ref{riemannian_model}).}
	\label{alg:1}
	\begin{algorithmic}[1]
		\STATE {\bfseries Input:} $K\geq 1$ and sequences $\{\eta_k\}_{k=1}^T$;\\
		\STATE {\bfseries Initialize:} $\bm{x}_0\in \mathcal\bm{{M}_r}$;\\
		\FOR{$k=1,\ldots,T$}
		\STATE $\nabla f(\bm{x}_k)=A^*(A\bm{x}_k-\bm{y})+\lambda \nabla D(\bm{x}_k)$;\
		\STATE $\xi_k=\text{P}_{T_{\bm{x}}\mathcal\bm{{M}_r}}(\nabla f(\bm{x}_k))$;\
		\STATE $\bm{x}_{k+1}=R(\bm{x}_k,-\eta_k\xi_k)$;\
		\ENDFOR
		\STATE {\bfseries Output:} $\bm{x}_{K}.$
	\end{algorithmic}
\end{algorithm}


\subsection{The Proposed Network}
Although the iterative scheme of Riemannian GD for the problem (\ref{riemannian_model}) was given, there are two intractable problems: both the hyper-parameters $\{\lambda,\eta_k\}$ and regularizer $D(\cdot)$ need to be selected empirically, which is tedious and uncertain. What's worse, the iterative solution often takes a long time, which limits its clinical application.

\begin{figure}[htbp]
	\centerline{\includegraphics[width=1.0\linewidth]{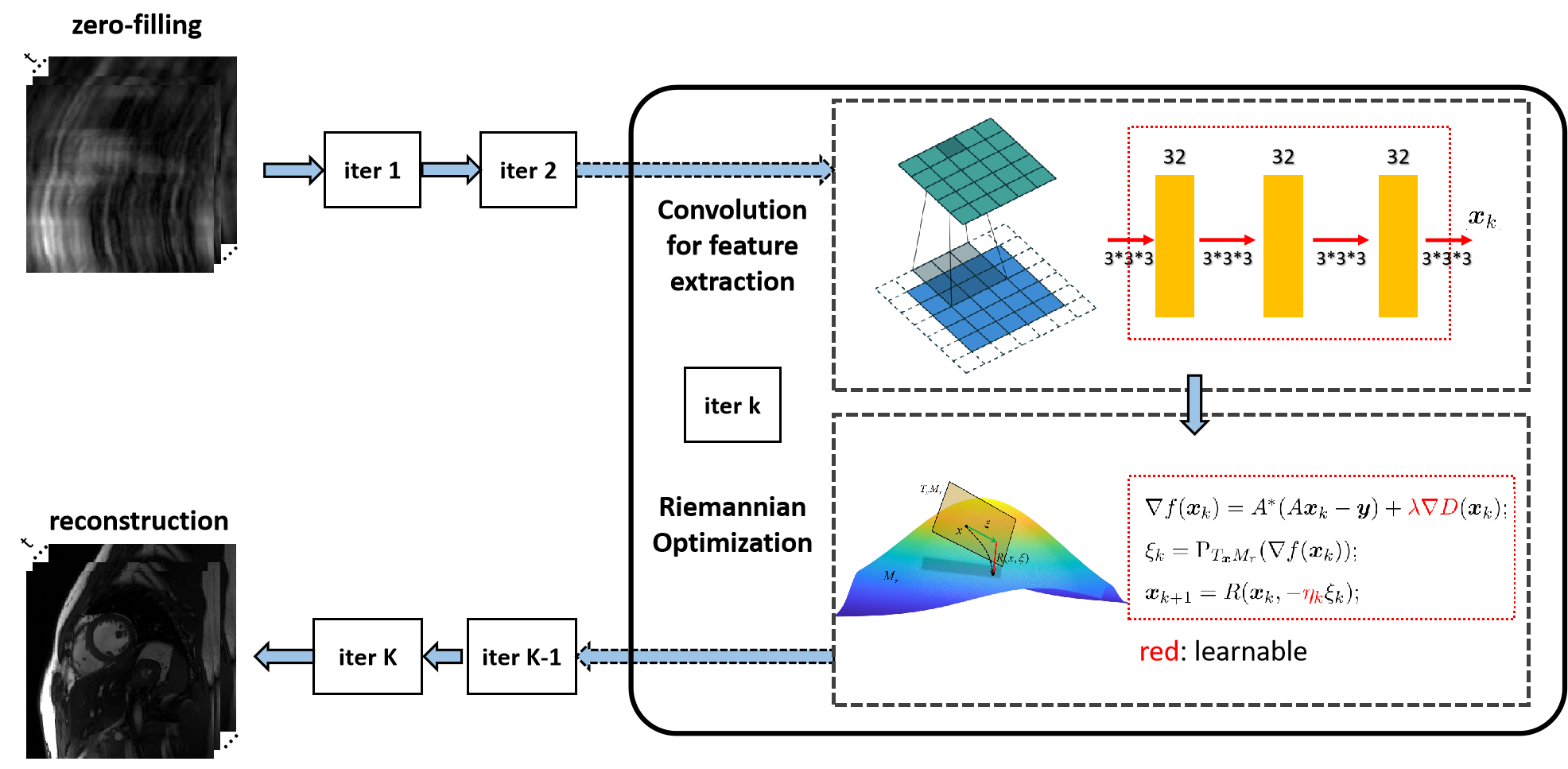}}
	\caption{The proposed Manifold-Net based on Riemannian optimization. Manifold-Net is composed of multiple cascaded neural network blocks. Each block contains convolution and Riemannian optimization.  \label{manifold_net}}
\end{figure}

To address the above two issues, we propose a deep Riemannian network, dubbed as Manifold-Net. Manifold-Net is composed of multiple cascaded neural network blocks depicted in Fig. \ref{manifold_net}. Each block contains a convolution module and a Riemannian optimization module. The convolution module is made up of three convolution layers for feature extraction. The Riemannian optimization module unrolls Algorithm \ref{alg:1} into a deep neural network. Specifically, the three procedures participate in the forward process of the network, the hyperparameters $\{\lambda,\eta_k\}$ and regularizer $D(\cdot)$ are set to be learnable by the network. Among them, the hyperparameters are defined as the learnable variables of the network, and the regularizer is learned by the convolutional neural networks. 

Our proposed deep Manifold-Net has the following advantages: 1) Every iteration of the Riemannian network always stays on the manifold $\mathcal\bm{{M}_r}$, and the calculations of gradient and iterative trajectory always follows the manifold structure itself. 2). The proposed network can learn all of the hyperparameters and transforms, which eliminates the complex and lengthy selection of parameters and transforms. 3). Once the optimal network parameters are learned, we can reconstruct images with good quality in seconds, as the network avoids the tedious iteration associated with traditional low-rank dynamic MRI. This work represents the first study unrolling the optimization on manifolds into neural networks.
\section{Results}
\subsection{Setup}
\subsubsection{Data acquisition}
The fully sampled cardiac cine data were collected from 30 healthy volunteers on a 3T scanner (MAGNETOM Trio, Siemens Healthcare, Erlangen, Germany) with a 20-channel receiver coil array. All in vivo experiments were approved by the Institutional Review Board (IRB) of Shenzhen Institutes of Advanced Technology, and informed consent was obtained from each volunteer. Relevant image parameters of our cine sequence included the following. For each subject, 10 to 13 short-axis slices were imaged with the retrospective electrocardiogram (ECG)-gated segmented bSSFP sequence during breath-hold. A total of 386 slices were collected. The following sequence parameters were used: FOV = $330\times330$ mm, acquisition matrix = $256\times256$, slice thickness = 6 mm, and TR/TE = 3.0 ms/1.5 ms. The acquired temporal resolution was 40.0 ms and reconstructed to produce 25 phases to cover the entire cardiac cycle. The raw multi-coil data of each frame were combined by an adaptive coil combine method \cite{Walsh2000Adaptive} to produce a single-coil complex-valued image.  We randomly selected images from 25 volunteers for training and the rest for testing. Deep learning typically requires a large amount of data for training \cite{lecun2015deep}. Therefore, we applied data augmentation using rigid transformation-shearing to enlarge the training pool. We sheared the dynamic images along the x, y, and t directions. The sheared size was $192\times192\times18$ ($x\times y\times t$), and the stride along the three directions was 25, 25, and 7. Finally, we obtained 800 2D-t cardiac MR data of size $192\times192\times18$ ($x\times y\times t$) for training and 118 data for testing. 

\begin{figure}[htbp]
	\centerline{\includegraphics[width=1.0\linewidth]{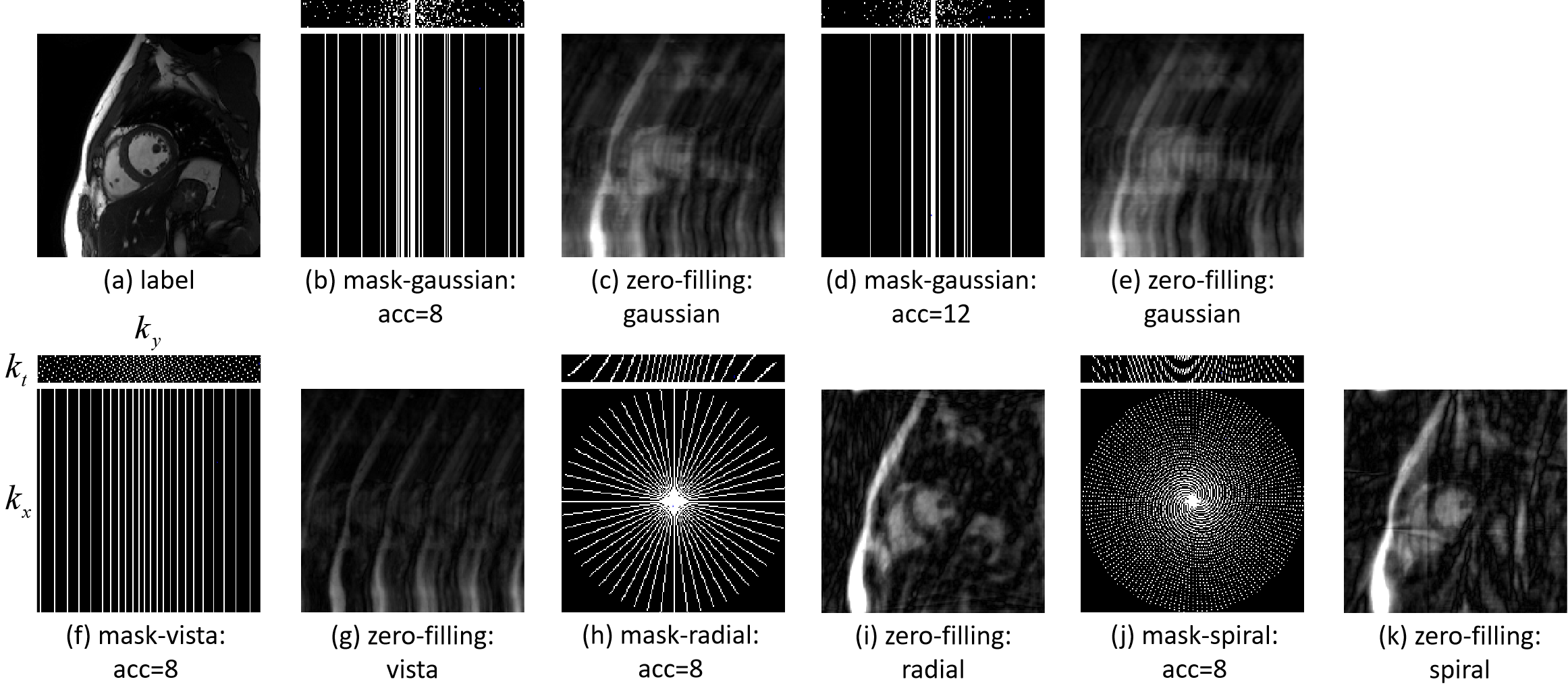}}
	\caption{The undersampling masks used in this work. (a) Label. (b) Gaussian random mask (8-fold). (c) Zero-filling image (Gaussian, 8-fold). (d) Gaussian random mask (12-fold). (e) Zero-filling image (Gaussian, 12-fold). (f) Vista mask (8-fold). (g) Zero-filling image (Vista, 8-fold). (h) Radial mask (8-fold). (i) Zero-filling image (Radial, 8-fold). (j) Spiral mask (8-fold). (k) Zero-filling image (Spiral, 8-fold). \label{masks}}
\end{figure}

Retrospective undersampling was performed to generate input/output pairs for network training. We fully sampled frequency encodes (along with $k_x$) and randomly undersampled phase encodes (along $k_y$) according to a zero-mean Gaussian variable density function \cite{jung2007improved} as shown in Fig \ref{masks}. Wherein four central phase encodes were ensured to be sampled.

\subsubsection{Model configuration}
To demonstrate the effectiveness and flexibility of the Manifold-Net in dynamic MR cine imaging, we compared the results of Manifold-Net with the single-coil version of a CS-based method (k-t SLR \cite{lingala2011accelerated}) and two state-of-the-art CNN-based methods (DC-CNN \cite{schlemper2017deep}, and CRNN \cite{qin2018convolutional}). We did not compare with MoDL-SToRM \cite{biswas2019dynamic} since the SToRM acquisition relied on navigator signals that were used to compute the manifold Laplacian matrix, while we acquired the data without a navigator and its source code is not publicly available. All comparison methods were executed according to the source code provided by the authors. For a fair comparison, all the methods mentioned in this paper were adjusted to their best performance.

We divided each data into two channels for network training, where the channels stored real and imaginary parts of the data. Therefore, the inputs of the network were undersampled k-space $\mathbb{C}^{2N_xN_yN_t}$, and the outputs were reconstruction images $\mathbb{C}^{2N_xN_yN_t}$. Manifold-Net has ten iterative steps; that is, $K=10$. The rank selection of a fixed-rank manifold is 13; that is $r=13$. The learned transforms, $\{D\}$, are different for each layer.  Each convolutional layer had 32 convolution kernels, and the size of each convolution kernel was $3 \times 3 \times 3$. He initialization \cite{he2015delving} was used to initialize the network weights. Rectifier linear units (ReLU) \cite{glorot2011deep} were selected as the nonlinear activation functions. The mini-batch size was 1. The exponential decay learning rate \cite{zeiler2012adadelta} was used in all CNN-based experiments with an initial learning rate of 0.001 and a decay of 0.95. All the models were trained by the Adam optimizer \cite{kingma2014adam} with parameters $\beta_1=0.9$, $\beta_2=0.999$, and $\epsilon=10^{-8}$ to minimize a mean square error (MSE) loss function. Code is available at \url{https://github.com/Keziwen/Manifold_Net}.

The models were implemented on an Ubuntu 16.04 LTS (64-bit) operating system equipped with an Intel Xeon Gold 5120 central processing unit (CPU) and an Nvidia RTX 8000 graphics processing unit (GPU, 48 GB memory) in the open framework TensorFlow \cite{abadi2016tensorflow} with CUDA and CUDNN support. The network training took approximately 36 hours within 50 epochs. 

\subsubsection{Performance evaluation}
For a quantitative evaluation, the MSE, peak-signal-to-noise ratio (PSNR), and structural similarity index (SSIM) \cite{wang2004image} were measured as follows:
\begin{equation}
\label{MSE}
\mathrm{MSE}=
||Ref-Rec||^2_2
\end{equation}
\begin{equation}
\label{PSNR}
\mathrm{PSNR}
= 20\log_{10}\frac{\max(Ref)\sqrt{N}}{||Ref-Rec||_2}
\end{equation}
\begin{equation}
\label{SSIM}
\mathrm{SSIM}
= \boldsymbol{l}(Ref, Rec)\cdot\boldsymbol{c}(Ref, Rec)\cdot\boldsymbol{s}(Ref, Rec)
\end{equation}
where $Rec$ is the reconstructed image, $Ref$ denotes the reference image, and $N$ is the total number of image pixels. The SSIM index is a multiplicative combination of the luminance term, the contrast term, and the structural term (details are shown in \cite{wang2004image}).

\subsection{The Reconstruction Performance of the Proposed Manifold-Net}
\begin{figure}[htbp]
	\centerline{\includegraphics[width=1.0\linewidth]{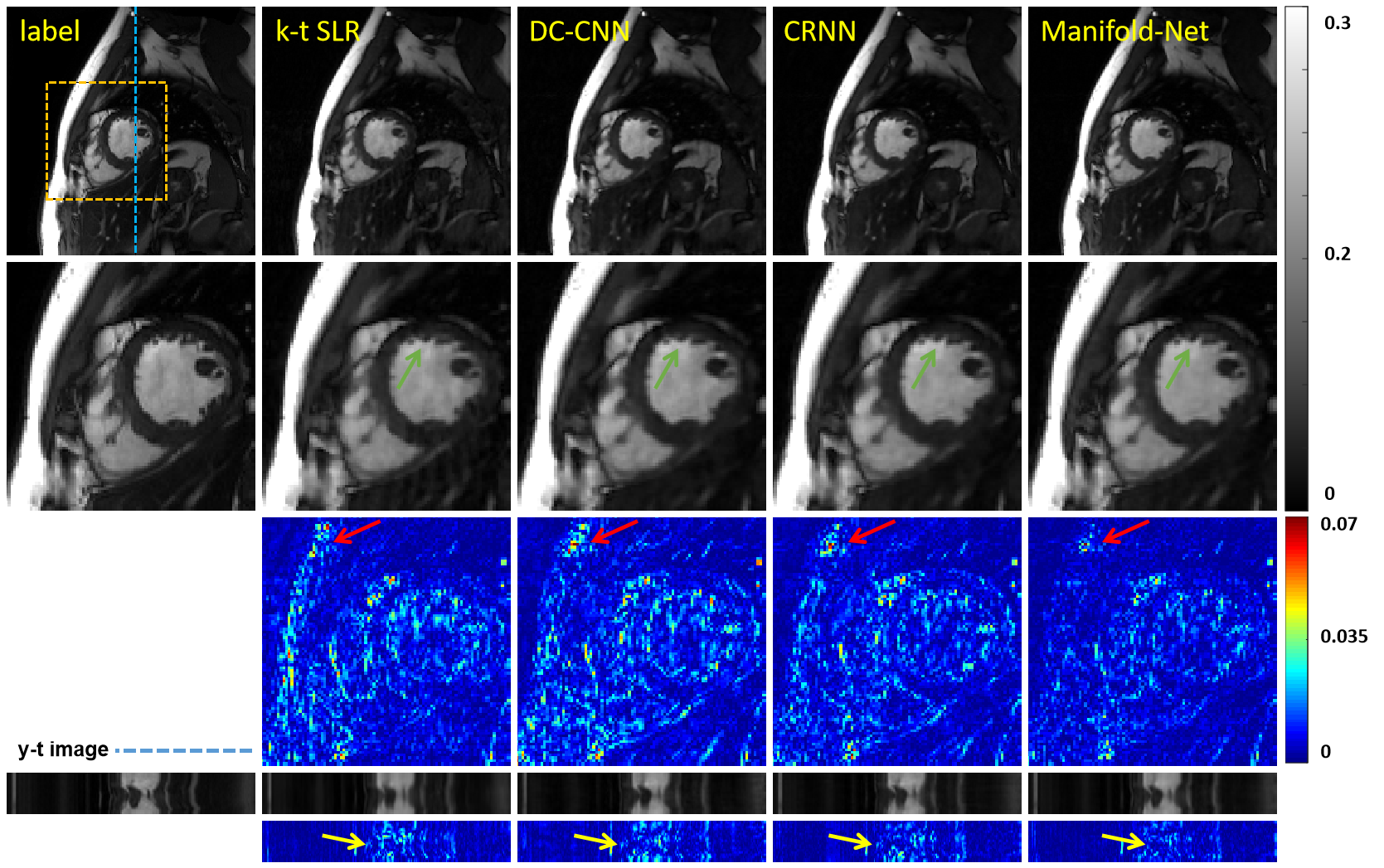}}
	\caption{The reconstruction results of the different methods (k-t SLR, DC-CNN, CRNN, and the proposed Manifold-Net) at 8-fold acceleration. The first row shows, from left to right, the ground truth and the reconstruction results of these methods. The second row shows the enlarged view of their respective heart regions framed by a yellow box. The third row shows the error map (display ranges [0, 0.07]). The y-t image (extraction of the 124th slice along the y and temporal dimensions, as marked by the blue dotted line) and the error of y-t image are also given for each signal to show the reconstruction performance in the temporal dimension. \label{results_acc_8}}
\end{figure}

To demonstrate the efficacy of the proposed Manifold-Net in the single-coil scenario, we compared it with a CS-LR method; namely, k-t SLR \cite{lingala2011accelerated}, and two sparse-based CNN methods, namely, DC-CNN \cite{schlemper2017deep}, and CRNN \cite{qin2018convolutional}. The reconstruction results of these methods at 8-fold acceleration are shown in Fig.\ref{results_acc_8}. The first row shows, from left to right, the ground truth and the reconstruction results of the methods as marked in the figure. The second row shows the enlarged view of the corresponding heart regions framed by a yellow box. The third row shows the error map (display ranges [0, 0.07]). The y-t image (extraction of the 124th slice along the y and temporal dimensions, as marked by the blue dotted line), and the error of the y-t image are also given for each signal to show the reconstruction performance in the temporal dimension. The reconstruction performance of the three deep learning-based methods (DC-CNN, CRNN, and Manifold-Net) is better than that of the traditional iterative method (k-t SLR), which can be clearly seen from the error maps. The comparison between the three deep learning methods shows that Manifold-Net is better than the other three methods in both detail retention and artifact removal (as shown by the green and red arrows). The y-t results also have consistent conclusions, as shown by the yellow arrows. The numbers of parameters for these network models are provided in Table \ref{quan_acc_8}. Manifold-Net has the minimum number of parameters, so it can be concluded that Manifold-Net gets the optimal reconstruction results due to the deep learning-based Riemannian optimization rather than a larger network capacity.

We also provide quantitative evaluations (MSE, PSNR, SSIM) in Table \ref{quan_acc_8}. Manifold-Net achieves optimal quantitative evaluations (MSE, PSNR, and SSIM). Both qualitative and quantitative results demonstrate that the proposed Manifold-Net can effectively explore the low-rank prior of dynamic data, thus improving the reconstruction performance. It is also proved that the deep learning-based optimization on manifolds is feasible.
\begin{table*}[!t]
	\renewcommand{\arraystretch}{1.1}
	\renewcommand\tabcolsep{1.5pt}
	\caption{\textcolor{black}{The average MSE, PSNR, SSIM of k-t SLR, DC-CNN, CRNN and Manifold-Net on the test dataset at 8-fold acceleration (mean$\pm$std).}}
	\label{quan_acc_8}
	\centering
	\textcolor{black}{\begin{tabular}{l|cccc}\hline\hline
			Methods&MSE(*e-5)&PSNR&SSIM(*e-2)&Parameters(*e+4)\\\hline
			k-t SLR&$8.22\pm3.04$&$41.14\pm1.57$&$95.11\pm0.88$&$/$\\
			DC-CNN&$7.43\pm2.33$&$41.48\pm1.30$&$96.22\pm0.76$&$43.2650$\\
			CRNN&$5.60\pm1.67$&$42.70\pm1.24$&$97.07\pm0.61$&$29.7794$\\
			Manifold-Net&$\bm{3.50\pm0.58}$&$\bm{44.62\pm0.77}$&$\bm{97.95\pm0.31}$&$\bm{11.2325}$\\
			\hline\hline 
	\end{tabular}}
\end{table*}
\section{Discussion}

\subsection{Higher Acceleration: 12-fold} 
\begin{figure*}[htbp]
	\centerline{\includegraphics[width=1.0\linewidth]{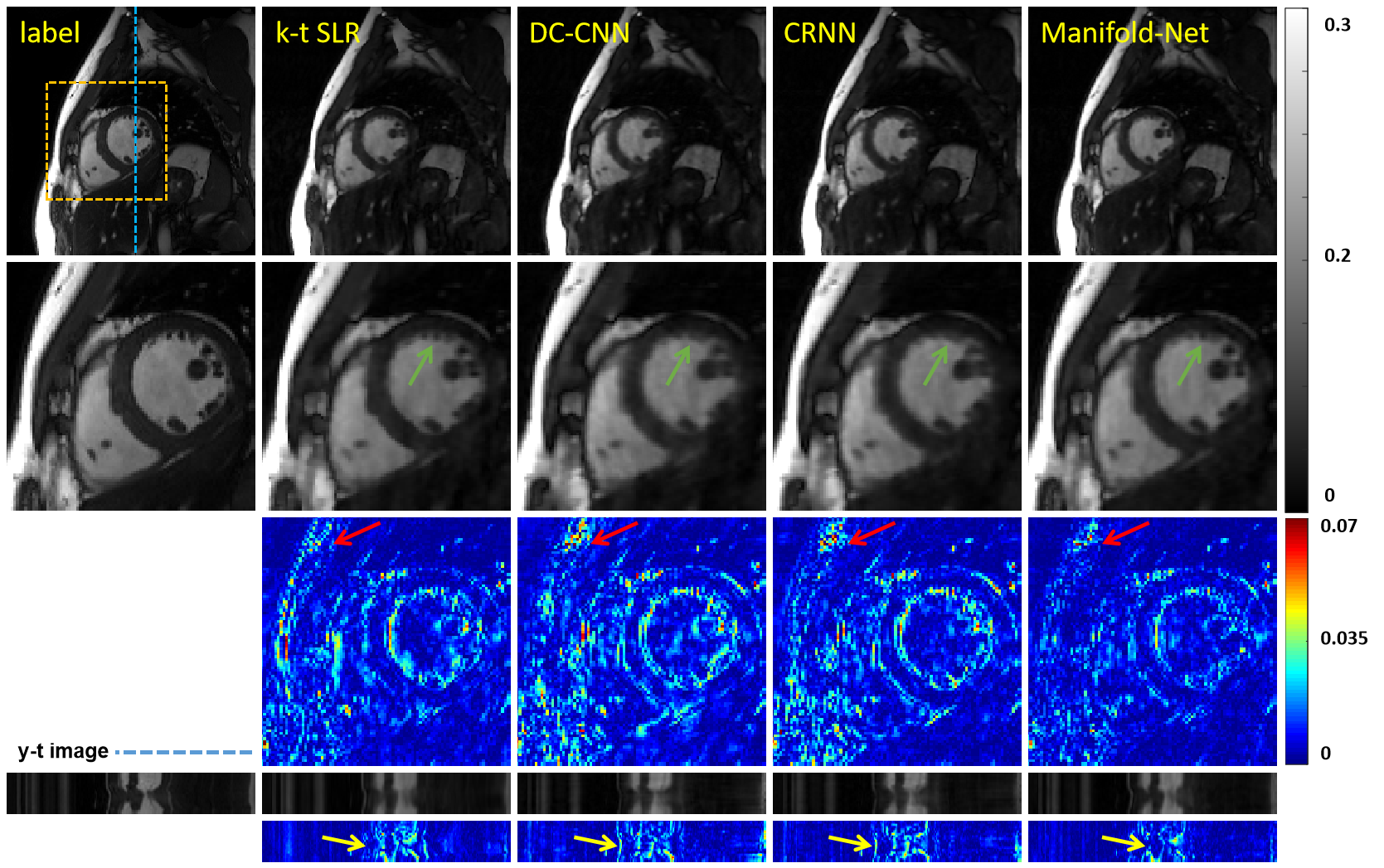}}
	\caption{The reconstruction results of the proposed Manifold-Net at  12-fold accelerations in the single-coil scenario. The first row shows, from left to right, the ground truth and the reconstruction results of these methods. The second row shows the enlarged views of their respective heart regions framed by a yellow box. The third row shows the error maps (display ranges [0, 0.07]). The y-t image (extraction of the 124th slice along the y and temporal dimensions, as marked by the blue dotted line) and the error of the y-t image are also given for each signal to show the reconstruction performance in the temporal dimension. \label{results_acc_12}}
\end{figure*}

The proposed method can explore the low-rank priors of dynamic signals on the manifold, which not only improves the reconstruction performance but also increases the acceleration rate because more expert knowledge is introduced into the optimization problem. We explore the reconstruction performance at higher accelerations in a single-coil scenario. The 12-fold accelerated reconstruction results can be found in Fig. \ref{results_acc_12}. Our proposed Manifold-Net still achieves superior reconstruction performance at 12-fold acceleration. Although the results are slightly vague, most of the details are well preserved. The quantitative indicators are provided in Table \ref{quan_acc_12}, which confirms that our proposed Manifold-Net still achieves excellent quantitative performance at higher accelerations.
\begin{table*}[!t]
	\renewcommand{\arraystretch}{1.1}
	\renewcommand\tabcolsep{1.5pt}
	\caption{\textcolor{black}{The average MSE, PSNR, SSIM of k-t SLR, DC-CNN, CRNN and Manifold-Net on the test dataset at 12-fold acceleration (mean$\pm$std).}}
	\label{quan_acc_12}
	\centering
	\textcolor{black}{\begin{tabular}{l|ccc}\hline\hline
			Methods&MSE(*e-5)&PSNR&SSIM(*e-2)\\\hline
			k-t SLR&$14.66\pm6.62$&$38.76\pm1.91$&$91.71\pm2.52$\\
			DC-CNN&$12.98\pm3.62$&$39.03\pm1.19$&$93.78\pm0.87$\\
			CRNN&$11.87\pm3.35$&$39.43\pm1.21$&$94.57\pm0.89$\\
			Manifold-Net&$\bm{6.46\pm0.97}$&$\bm{41.95\pm0.67}$&$\bm{96.37\pm0.40}$\\
			\hline\hline 
	\end{tabular}}
\end{table*}

\subsection{The Sensitivity to Different Undersampling Masks}
\begin{figure*}[htbp]
	\centerline{\includegraphics[width=1.0\linewidth]{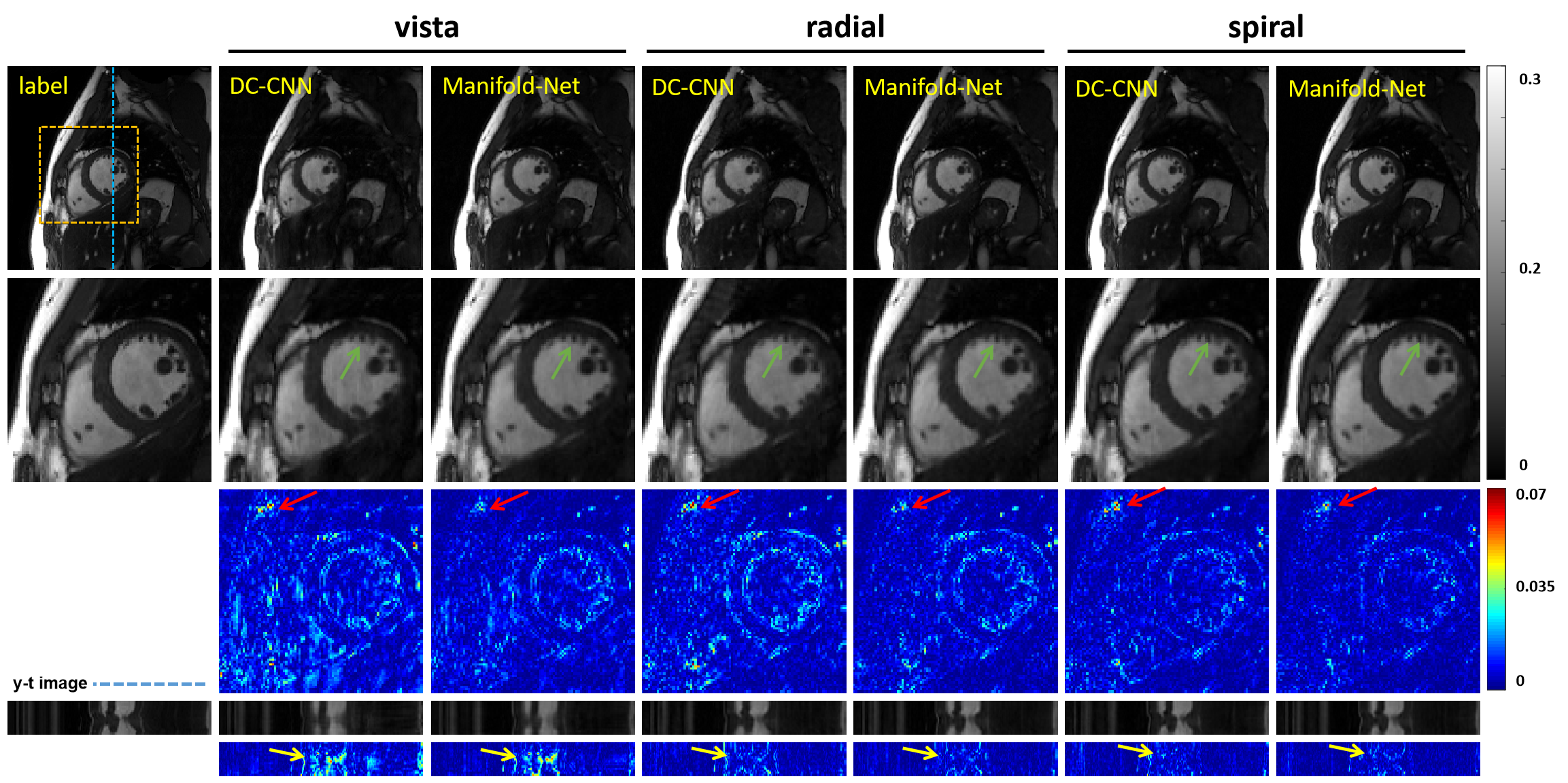}}
	\caption{The reconstruction results of the proposed Manifold-Net under different undersampling masks (radial, spiral, and VISTA) at 8-fold  acceleration. \label{results_diff_mask}}
\end{figure*}

The proposed Manifold-Net has good reconstruction performance for different undersampling masks. In this section, as a proof-of-concept, we explored the results of Manifold-Net trained under different masks (radial \cite{feng2016xd}, spiral \cite{pruessmann2001advances}, and VISTA \cite{ahmad2015variable}) at 8-fold acceleration. The reconstruction results under different undersampling masks can be found in Fig. \ref{results_diff_mask}. Compared with DC-CNN \cite{schlemper2017deep}, the proposed Manifold-Net achieves better reconstruction results regardless of the mask, as shown by the red arrow. Especially under the VISTA undersampling mask, the reconstruction results of DC-CNN-Net are significantly poorer than those of radial and spiral methods. However, Manifold-Net still maintains good reconstruction results. The quantitative indicators, shown in Table \ref{quan_diff_mask}, confirm that our proposed Manifold-Net achieves better quantitative performance under each undersampling mask.
\begin{table}[!t]
	\renewcommand{\arraystretch}{1.1}
	\renewcommand\tabcolsep{1.5pt}
	\caption{\textcolor{black}{The average MSE, PSNR and SSIM of DC-CNN and Manifold-Net under different undersampling masks at 8-fold acceleration on the test dataset (mean$\pm$std).}}
	\label{quan_diff_mask}
	\centering
	\textcolor{black}{\begin{tabular}{ll|ccc}\hline\hline
			&Methods&MSE(*e-5)&PSNR&SSIM(*e-2)\\\hline
			&DC-CNN&$12.02\pm3.83$&$39.43\pm1.43$&$92.71\pm1.32$\\
			Vista$ $&Manifold-Net&$\bm{5.64\pm0.84}$&$\bm{42.53\pm0.63}$&$\bm{96.34\pm0.45}$
			\\
			\hline
			&DC-CNN&$5.46\pm1.91$&$42.88\pm1.49$&$96.84\pm0.83$\\
			Radial$ $&Manifold-Net&$\bm{2.58\pm0.42}$&$\bm{45.94\pm0.73}$&$\bm{98.38\pm0.24}$
			\\
			\hline
			&DC-CNN&$4.05\pm1.41$&$44.17\pm1.46$&$97.50\pm0.68$\\
			Spiral$ $&Manifold-Net&$\bm{2.15\pm0.32}$&$\bm{46.73\pm0.67}$&$\bm{98.60\pm0.21}$
			\\
			\hline\hline 
	\end{tabular}}
\end{table}

\subsection{The Rank Selection: $r$}
\begin{figure*}[htbp]
	\centerline{\includegraphics[width=1.0\linewidth]{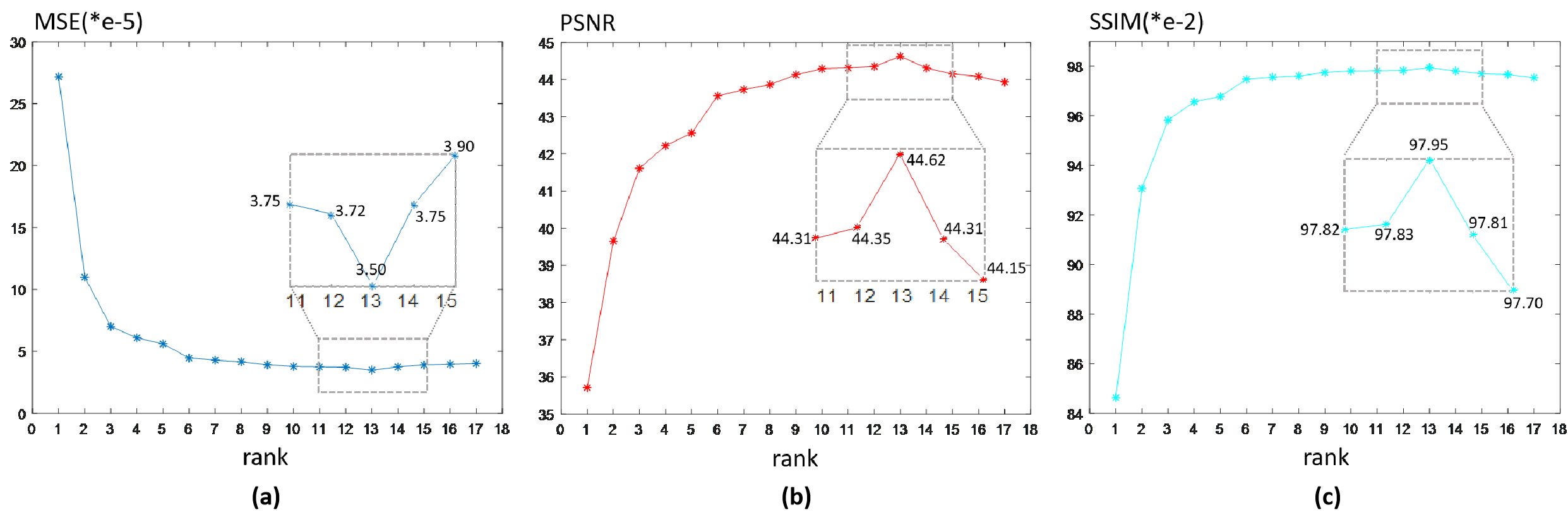}}
	\caption{The average MSE, PSNR, SSIM of  Manifold-Net under different ranks on the test dataset at 8-fold acceleration. (a) MSE. (b) PSNR. (c) SSIM. \label{rank_plot}}
\end{figure*}

We designed a fixed rank manifold, $\mathcal\bm{{M}_r}$, to describe the temporal redundancy of dynamic signals. The selection of rank greatly influences the reconstruction performance, which is discussed in this section. The quantitative indicators under different ranks are given in Fig. \ref{rank_plot}. The best reconstruction performance is achieved when the rank $r$ is equal to 13. As the rank gets smaller, the reconstruction gets worse. This indicates that when the rank is 13, the designed manifold can well describe the temporal redundancy of the dynamic signals.

\subsection{The Limitations of the Proposed Manifold-Net}
Although the Manifold-Net achieves improved reconstruction results, it still has the following limitations: 1) Fixed-rank manifolds need to be characterized in advance, especially the selection of rank $r$. If the rank selection is not good, the reconstruction performance may be poor. 2) MRI is collected in multiple coils. Still, this paper only discusses the results in the single-coil scenario, and the effectiveness of the multi-coil version of Manifold-Net remains to be verified. 3) Due to the application of HOSVD in the Riemannian optimization, the reconstruction time reached 5-6s for the entire 18 frames of dynamic signals. An SVD-free strategy \cite{huang2020exponential} will be explored in future work.
\section{Conclusions}

This paper develops a deep learning method on a nonlinear manifold to explore the temporal redundancy of dynamic signals to reconstruct cardiac MRI data from highly undersampled measurements. Every iteration of the manifold network always stays on the designed manifold, and the calculations of gradient and iterative trajectory always follow the manifold structure itself. Experimental results at high accelerations demonstrate that the proposed method can obtain improved reconstruction compared with a compressed sensing (CS) method k-t SLR and two state-of-the-art deep learning-based methods, DC-CNN and CRNN.
This work represents the first study unrolling the optimization on manifolds into neural networks. Specifically, the designed low-rank manifold provides a new technical route for applying low-rank priors in dynamic MR imaging.

\section{Acknowledgments}

This work was supported in part by the National Key R$\&$D Program of China (2020YFA0712202, 2017YFC0108802 and 2017YFC0112903); National Natural Science Foundation of China (61771463, 81830056, U1805261, 81971611, 61871373, 81729003, 81901736); Natural Science Foundation of Guangdong Province (2018A0303130132); Shenzhen Key Laboratory of Ultrasound Imaging and Therapy (ZDSYS20180206180631473); Shenzhen Peacock Plan Team Program (KQTD20180413181834876); Innovation and Technology Commission of the government of Hong Kong SAR (MRP/001/18X); Strategic Priority Research Program of Chinese Academy of Sciences (XDB25000000); Key Laboratory for Magnetic Resonance and Multimodality Imaging of Guangdong Province (2020B1212060051)


\bibliographystyle{unsrt}
\bibliography{manifoldnet} 


\end{document}